\documentclass{kapproc}
\usepackage{amssymb}

\usepackage{graphicx}

\normallatexbib

\begin{document}

\articletitle[Nuclear Spin Relaxation]{NUCLEAR SPIN RELAXATION
\\ AND INCOMMENSURATE MAGNETISM IN DOPED CUPRATES}

\author{L.\ P.\ Gor'kov,\thanks{
E-mail: gorkov@magnet.fsu.edu}}
\affil{National High Magnetic Field Laboratory, Florida State
University,
Tallahassee,\\ FL 32310, USA\\
and L.D.Landau
Institute for Theoretical Physics, 142432 Chernogolovka, Russia}
\author{G.\ B.\ Teitel'baum,\thanks{
E-mail: grteit@dionis.kfti.knc.ru}}
\affil{E.K.Zavoiskii Institute for Technical Physics of the RAS,
420029 Kazan, Russia}

\date{\today }

\begin{abstract}
Existing data on $^{63}$Cu-nuclear spin relaxation reveal two
independent relaxation processes: the one that is temperature
independent we link to incommensurate peaks seen by neutrons,
while the ''universal'' temperature dependent contribution
coincides with $1/^{63}T_{1}(T)$ for two-chain YBCO 124. We argue
that this new result substitutes for a ''pseudogap'' regime in a
broad class of high-$T_{c}$ cuprates and stems from the 1st order
phase transition that starts well above the superconductivity
$T_{c}$ but becomes frustrated because of broken electroneutrality
in the CuO$_{2}$ plane.
\end{abstract}








%

\section{Introduction}
One of the most intriguing normal properties of the high-$T_{c}$
(HT$_{c}$) cuprates is the so called ''pseudogap'' (PG)
phenomenon. It is commonly presented in the $(T,x)$ plane as a
line that starts from rather high temperatures (at small $x$) and
reaches the superconductivity (SC) $T_{c}$ ''dome'' below at or
above optimal $x\sim 0.16$. In a broad sense $x$ means the hole
concentration in the CuO$_{2}$- plane, but more often than not one
refers to properties of the Sr-doped La$_{2-x}$Sr$_{x}$CuO$_{4}$ \
The PG feature was seen in numerous experiments (NMR, tunneling
spectra, resistivity etc.; see for example reviews \cite{14,15}).
It has been stressed \cite{27} that the PG temperature is not
defined unambiguously.

A widespread view is that the feature comes from some crossover in
the electronic density of states (DOS). The main result of the
present paper is that after a proper re-arrangement of the
experimental data no PG feature exists in the $^{63}$Cu nuclear
spin relaxation time behaviour. Instead, the data show two
independent parallel relaxation mechanisms: a temperature
independent one that we attribute to stripes caused by the
presence of external dopants and an ''universal'' temperature
dependent term which turns out to be exactly the same as in the
stoichiometric compound YBCO 124.

\section{The experimental results and discussion}
We attempt below to put the results in the context of a phase separation %
\cite{1}. The decomposition of $1/^{63}T_{1}(T,x)$ into two terms,
as it will be discussed below in more details, manifests itself in
a broad temperature interval above $T_{c}$. It is limited from
above by a $T^{\ast }$ that depends on the concentration, $x$. We
consider $T^{\ast }$ defined in this way as a temperature of a 1st
order phase transition, which, however, cannot complete itself in
spatial coexistence of two phases because of the electroneutrality
condition \cite{17}. It was already argued in \cite{1} that such a
frustrated 1st order phase transition may actually bear a
dynamical character. The fact that a single resonant frequency for
the $^{63}$Cu nuclear spin is observed in the NMR experiments,
confirms this suggestion. Although in what follows, we use the
notions of the lattice model \cite{1,17}, even purely electronic
models \cite{2,3,4,5} for cuprates may reveal a tendency to phase
separation.

The basic assumption in \cite{1,17} are the following. At large
enough doping holes move between coppers and oxygens. Spins in the
system are $d^{9} $-holes trapped to the Cu-sites at the expense
of local lattice distorsions. Elastic attractive interactions
between these distorsions give rise  to a lattice driven
frustrated
transition below some $T^{\ast }$. Exchange interactions, as in the parent La$%
_{2}$CuO$_{4}$, tend to organize the Cu-spins in the antiferromagnetic (AF)
sub-phases. Excess charge of the dopants' ions in AF regions must be
compensated by accumulation of holes in ''metallic'' regions.

We now turn to experimental data. In what follows we address only $%
1/^{63}T_{1}$ behaviour because for cuprates AF fluctuations prevail over
the Korringa mechanisms.

\begin{figure}[ht]
\centering{\includegraphics[width=8cm]{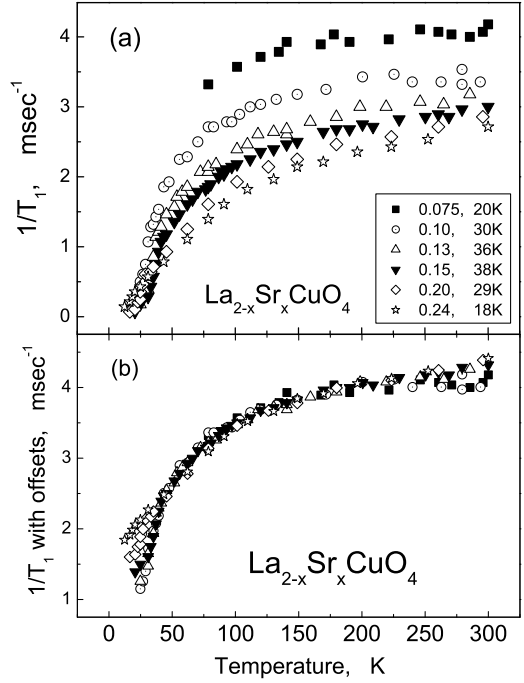}} \caption{The
temperature dependence of $1/^{63}T_{1}(x)$ for LSCO: a) the
plots for different $x$ and $T_{c}$ (see inset) are taken from \protect\cite%
{39}, at higher temperatures all of them converge to the same value of 2.7
msec$^{-1}$ \protect\cite{Imai} ; b) the same dependencies collapsing to the
single curve after the corresponding vertical offsets.}
\label{Fig1}
\end{figure}

In Fig. 1a we collected data on $1/^{63}T_{1}$ in LSCO from \cite{39}. Note
the following: 1) according to \cite{Imai} $1/^{63}T_{1}(T)$ at higher
temperatures tends to 2.7 msec$^{-1}$ \textit{for all} Sr concentrations, in
spite of considerable spread seen in Fig. 1a. Beginning of deviation from
that value could serve us as a definition of $T^{\ast }(x)$; 2) note that
dissipation $1/^{63}T_{1}$\ monotonically decreases from small $x$ to 0.24;
3) after an appropriate \textit{vertical} offset all curves in Fig. 1a
collapse onto the $T$ dependence of $1/^{63}T_{1}$ for the ``optimal'' $%
x=0.15$ above 50 K (Fig. 1b). We have checked that last tendency
works well
for YBCO (6.5) doped with Ca i.e., the data for different $z$ in Y$_{1-z}$Ca$%
_{z}$Ba$_{2}$ Cu$_{3}$ O$_{6.5}$ \cite{Singer} may be put  on the top of
each other after proper offsets.

\sloppy
This prompts us to verify whether same ''off-settings'' of
the $1/^{63}T_{1}$
data apply to a broader group of materials. The stoichiometric YBa$_{2}$Cu$%
_{4}$O$_{8}$ possesses no structural or defect disorder and we
adjust all data to the $1/^{63}T_{1}$ behaviour for this material
\cite{27}. Fig. 2 shows that after a vertical shift in
$1/^{63}T_{1}$ all the materials indeed follow the same
''universal'' temperature dependence above their $T_{c}$ and below
300 K. In other words, in this temperature range the nuclear spin
relaxation in these cuprates is a sum of contributions from two
parallel processes:
\begin{equation}
1/^{63}T_{1}=1/^{63}\overline{T}_{1}(x)+1/^{63}\widetilde{T}_{1}(T)
\end{equation}%
In eq. (1) $1/^{63}\overline{T}_{1}(x)$ depends on a material and a degree
of disorder $(x)$, but does not depend on temperature, while $1/^{63}%
\widetilde{T}_{1}(T)$, depends only on temperature, is the same for all
these compounds and coincides with the $1/^{63}T_{1}$ for the two chains
YBCO 124 above its $T_{c}$=62 K.

\begin{figure}[ht]
\centering{\includegraphics[width=8cm]{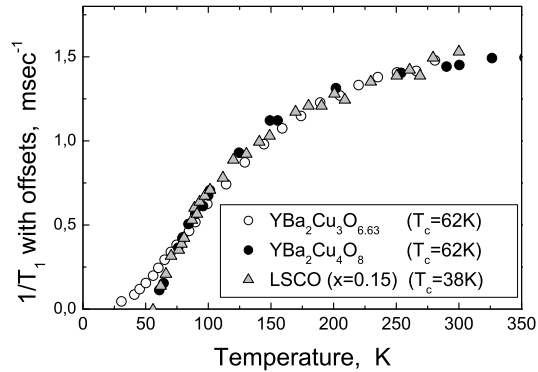}} \caption{The
temperature dependence of $1/^{63}T_{1}$ for different
compounds: the $1/^{63}T_{1}$ for YBCO (123) %
\protect\cite{26} overlayed with that for LSCO \protect\cite{39}
and YBCO (124) \protect\cite{ 27}.} \label{Fig2}
\end{figure}

Thus, to some surprise the only ''pseudogap'' feature in the NMR
data that may be discerned in Fig.2 is the one for YBCO 124: a
change in the temperature regime between 130 and 180 K. It would
be tempting to take again this feature as a mark of the PS regime
taking place now in the stoichiometric material where doping most
definitely comes about as a spill-over of carriers from the
CuO-chains into CuO$_{2}$ planes. It is also natural to think that
the number of the transferred carriers is not small: actually the
low temperature Hall effect measurements \cite{31,33,34} show a
rapid increase in the number of carriers (i.e. Fermi surface size)
up to one hole per unit cell
even in the single layer material like LSCO, at the optimal doping $x\sim$ %
0.15. Recall, however, that little is known for the
''homogeneous'' phase (i.e. above $T^{\ast }(x)$). Properties of
both YBCO 124 and the optimally doped LSCO (see \cite{14} for
review) are unusual and best described in a
very broad temperature interval in terms of the ''marginal'' Fermi liquid %
\cite{Varma}. We have not found a reliable experiment to define $T^{\ast }$
for these compounds and therefore leave the origin of the $1/^{63}\widetilde{%
T}_{1}(T)$-term for further discussions.

The decomposition (1) into two parallel dissipation processes show that
usual definitions of $T^{\ast }$ \cite{Schmallian} have no grounds. In
Fig.1a the LSCO data with $x<0.15$ are spread even above 250 K. As a rough
estimate for $T^{\ast }$, it is much higher than the SC onset temperature.

Fig. 3 presents the dependence on $x$ for $1/^{63}\overline{T}_{1}$ in La$%
_{2-x}$Sr$_{x}$CuO$_{4}.$ The inset provides the ''offsets'' (i.e. $1/^{63}%
\overline{T}_{1}$ terms) for other materials. We return to
discussion of Fig.3 later.

\begin{figure}[ht]
\centering{\includegraphics[width= 8 cm]{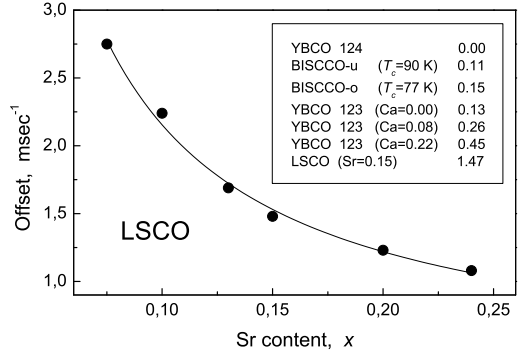}} \caption{The
offset $\ 1/^{63}\overline{T}_{1}(x)$ vs Sr content $x$ for LSCO
(relative to that for YBCO 124), line is a guide for eyes. Inset:
the offsets for some other compounds (data for underdoped (u) and
overdoped (o) BISCCO 2212 deduced from \protect\cite{Walstedt}; to
compare BISCCO with LSCO and YBCO materials the hyperfine
constants have to be properly adjusted).} \label{Fig3}
\end{figure}

The observation that is central for the following is that in all
the materials with non-zero $1/^{63}\overline{T}_{1}$ (see the
inset in Fig.3) incommensurate (IC) peaks have been observed in
neutron scattering \cite{G}. Peaks are close to the $[\pi ,\pi ]$
-- point: at $[\pi (1\pm \delta ),\pi ]$ and $[\pi ,\pi (1\pm
\delta )]$ \cite{9}. We will now look for the connection between
these two phenomena.

We first make an attempt to agree on a semi-quantitative level the observed
IC magnetic peaks in La$_{2-x}$Sr$_{x}$CuO$_{4}$ with the values of the
first term in eq. (1). We concentrate on La$_{1.86}$Sr$_{0.14}$CuO$_{4}$ for
which the most detailed data are available \cite{Aeppli}.

With the notation from \cite{Zha}
\begin{equation}
1/T_{1}=\frac{k_{B}T}{2\mu _{B}^{2}\hbar ^{2}\omega }\sum\limits_{i}F(Q_{i})%
\int \frac{d^{2}q}{(2\pi )^{2}}\chi ^{\prime \prime }(q,\omega \rightarrow 0)
\end{equation}%
where $Q_{i}$ stands for one peak, \ hyperfine ''tensor'' $\ F(Q)=\{A_{\perp
}+2B[\cos (Q_{x})+\cos (Q_{y})]\}^{2}$ and for $\chi ^{\prime \prime
}(q,\omega \rightarrow 0)$ we take near single peak, say $[\pi (1-\delta
),\pi ]$
\begin{equation}
\chi ^{\prime \prime }(q,\omega )=\frac{\chi _{peak}^{\prime \prime
}(T)\omega }{\left[ 1+(x\xi _{x})^{2}+(y\xi _{y})^{2}\right] ^{2}}
\end{equation}%
where $(x,y)=(q_{x}-\pi (1-\delta );q_{y}-\pi )$ and $\xi _{x}$ and $\xi
_{y} $ are the correlation lengths in the two proper directions. After
integration the contribution from stripes with $q$ along the $x$-direction
is
\begin{equation}
1/^{63}T_{1}=\frac{k_{B}T}{\pi \mu _{B}^{2}\hbar \xi _{x}\xi _{y}}\{A_{\perp
}-2B[\cos (\pi \delta )+1]\}^{2}\chi _{peak}^{\prime \prime }
\end{equation}%
Experimentally \cite{Aeppli} $\chi _{peak}^{\prime \prime }(T)\propto T^{-2}$
and for $x=0.14$ $\ \ \ \delta =0.245\sim $1/4.\ Assuming the $T^{-1}$
dependence \cite{Aeppli} only for the one of $\xi $'s, $\xi _{x}$ and using\
for $A_{\perp }$ and $B$ the known values \cite{Zha} one obtains: $%
1/^{63}T_{1}=(4/\xi _{y})$ msec$^{-1}$. With the AF correlation
length $\xi _{y}\sim $ 4 this is the correct order of magnitude.

The descending dependence of the offset (Fig.3) agrees qualitatively with
the behavior of $\delta (x)$ \cite{Yamada} in eq.(4). For a quantitative
description one need to know the $x$-dependence for $\chi _{peak}^{\prime
\prime }(T)$. Such data in the absolute units are absent yet except \cite%
{Aeppli}. Another fact that may underlie this behaviour is that
with the \textit{x}-increase buckling in the CuO$_{2}$-planes is
known to decrease
diminishing pinning effects and making the local symmetry of the CuO$_{2}$%
-unit same as in other materials from the class with small offset in Fig.3.
Also, the system grows more metallic with a high holes` content \cite%
{31,33,34}.

Next \ comes the question concerning the origin and the role played by IC
peaks and the physics of fluctuations related to them.

Discovery of IC spin fluctuations presented a challenge for explaining the
NMR results for the oxygen spin relaxation times: hyperfine field ''leaks''
originated by the AF incommensurate fluctuations, would considerably
increase the oxygen's relaxation rates, but this was not seen
experimentally. Slichter (see in Ref.\cite{Barzykin}) interpreted these
contradictions in terms of ''discommensurations'': a periodic array of
soliton-like walls separating regions with a short-range AF order. Unlike
neutrons, the NMR as a local probe, does not feel the overall periodicity.

Existence of stripes looks just natural in terms of a static phase
separation. At doping the system (LSCO) must screen the excess charge (Sr$%
^{2+}$-ions) in AF regions. Therefore stripes of the AF ordered phase must
alternate with ''metallic'' domain walls. The stripe arrangement by itself
is nothing but an optimization of the competing Coulomb and lattice forces %
\cite{3}. (The phenomena is well known in physics of surface.)

Stripes in a dynamical regime need better understanding. For instance, often
the IC peaks are seen by neutrons only at low enough temperatures or for
large energy transfer at an inelastic scattering \cite{G}. At low
temperatures stripes may form a long-range order even in LSCO (at smaller $x$%
, \cite{Fujita}), breaking the symmetry of the ground state. A better
example of the ''pinned'' stripe order is given by Nd (or Eu)-doped LSCO %
\cite{Crawford, Tranquada1, Tranquada2, Tranquada3}. (La$_{1.6-x}$Nd$_{0.4}$Sr$%
_{x}$CuO$_{4}$ reproduces all features of
La$_{1-x}$Sr$_{x}$CuO$_{4}$ , \ including SC and same positions
for IC peaks at given $x$.) The transition into the ordered stripe
phase is driven by appearance first of the lattice/charge peaks
\cite{Tranquada2}. At finite temperatures stripes could be viewed
as a new type of excitations above the ordered state.

On the other hand in LSCO itself ''stripes'' are seen through the
inelastic scattering processes for arbitrary low energy transfer
even at high temperatures 100-300 K \cite{Aeppli}. while the
ordered IC ground state sets in only well below, at about $T \sim$
30K \cite{Fujita}. \ This example provides the argument against
treating ''stripes'' as ''excitations'': at so high a temperature
the underlying ''long-range'' IC ground state would be already
melted. Therefore the two-phase description seems to be closer to
reality meaning that in the dynamical regime the AF regions get
coupled via Coulomb forces with the ''metallic'' layer. Note that
with the further x-increase \ $\delta (x)$ increases as well and
saturates making it meaningless to speak in terms of a strictly
''monolayer'' wall already above $x \sim$ 0.14, where $\delta
\approx 1/4$ \cite{Aeppli}. (Note the difference in notations for
IC peaks: ($\delta $ from \cite{Aeppli} equals $2\epsilon $ from
\cite{Tranquada2} equals $2\delta $ from \cite{Yamada}). Fig.3
demonstrates same tendency to increase the share of the
''metallic'' fraction with increase of Sr-concentration:
$1/^{63}\overline{T}_{1}(x)$ continues to drop with $x$ above 1/8.

Coexistence of a SC and the IC AF phases at low temperatures was confirmed
recently by the neutron diffraction experiments \cite{Lake} for La$_{2-x}$Sr$%
_{x}$CuO$_{4}$ $(x=0.10)$ in the vortex state. (The coexistence of SC and AF
formations was found also from the $\mu $SR spectra \cite{Niedermeier}). The
way of the ''coexistence'' of SC and the stripe order in the same sample
remains unresolved: one view treats the new stripe symmetry as a
superstructure superimposed on the Fermi surface that changes the energy
spectrum like any SDW/CDW can do it (e.g. \cite{Salcola}). Another plausible
alternative would be a spatially inhomogeneous coexistence of the
nonsuperconducting IC AF phase and a ''metallic'' phase with strong
fluctuations.

\section{Summary}
We have found that in a temperature interval above
$T_{c}$ and below some $T^{\ast}\sim $300 K the nuclear spin
relaxation for a broad class of cuprates comes from two
independent mechanisms: relaxation on the``stripe``-like
excitations that leads to a temperature independent contribution
to $1/^{63}T_{1}$ and an ``universal'' temperature dependent term.
For La$_{1.86}$Sr$_{0.14}$CuO$_{4}$ we obtained a correct
quantitative estimate for the value of the first term. We
concluded from eq.(1) and Fig.3 that ''stripes'' always come about
with external doping and may be pinned by structural defects. We
argue that the whole pattern fits well the notion of the dynamical
PS into coexisting metallic and IC magnetic phases.
Experimentally, it seems that with the temperature decrease
dynamical PS acquires the static character with the IC symmetry
breaking for AF phase dictated by the competition between the
lattice and the Coulomb forces. The form of coexistence of the IC
magnetism with SC below $T_{c}$ remains not understood as well as
behaviour of stoichiometric cuprates.

\section{Acknowledgements}
The work of L.P.G. was supported by the NHMFL through NSF
cooperative agreement DMR-9527035 and the State of Florida, that
of G.B.T. through the RFFR Grant N 01-02-17533.

\end{document}